# PROTON DIFFUSIVITY IN THE BaZr$_{0.9}$Y$_{0.1}$O$_{3-}$ PROTON CONDUCTOR

ARTUR BRAUN[1*], SOPHIE DUVAL[1], PETER RIED[1], JAN EMBS[2,3], FANNI JURANYI[2]
THIERRY STRÄSSLE[2], ULRICH STIMMING[4], ROLF HEMPELMANN[3]
PETER HOLTAPPELS[1], AND THOMAS GRAULE[1]

[1]*Laboratory for High Performance Ceramics*
*Empa - Swiss Federal Laboratories for Materials Testing & Research*
*CH-8600 Dübendorf, Switzerland*

[2]*Laboratory for Neutron Scattering*
*ETH Zurich & Paul Scherrer Institut, CH-5232 Villigen, Switzerland*

[3]*Physical Chemistry, Saarland University*
*D-66041 Saarbrücken, Germany*

[4]*Department of Physics E19*
*Technische Universität München, D - 85748 Garching, Germany*

\* Corresponding Author:
Phone: +41 44 823 4850
Fax: +41 44 823 4150
e-mail: artur.braun@alumni.ethz.ch

Abstract

The thermally activated proton diffusion in BaZr$_{0.9}$Y$_{0.1}$O$_{3-\delta}$ was studied with electrochemical impedance spectroscopy (IS) and quasi-elastic neutron scattering (QENS) in the temperature range from 300 K to 900 K. The diffusivities for the bulk material and the grain boundaries as obtained by IS obey an Arrhenius law with activation energies of 0.46 eV and 1.21 eV, respectively. The activation energies obtained by IS for the bulk are 0.26 eV above 700 K and 0.46 eV, below 700 K. The total diffusivity as obtained by IS is by one order of magnitude lower than the microscopic diffusivity as obtained by QENS. The activation energies obtained by QENS are 0.13 eV above 700 K and 0.04 eV, below 700 K. At about 700 K, the diffusion constants for IS and QENS have a remarkable crossover, suggesting two processes with different activation energies.

# 1. Introduction

Proton conductors (PC) are appealing to solid state electrochemists because they are compatible with aqueous, moist environments, and fit into a hydrogen economy. The fact that PC operate at temperatures as low as 700 K - 800 K, much lower than oxygen ion conductors (1000 K - 1300 K), makes them promising candidates for electrolytes in solid oxide fuel cell and electrolysers. Under ambient conditions, many oxide ceramics may absorb water molecules which then may dissociate [1] according to the following chemical reaction:

$$H_2O(g) + V_{\ddot{O}} + O_O^x \Leftrightarrow 2OH_O^{\cdot} \tag{1}$$

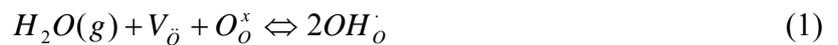

Proton diffusion is believed to be a two-step process, where, at moderate temperatures protons rotate locally around a lattice anion with very low activation energy in the order of few meV. At higher temperature, the proton may overcome the activation barrier of 0.1 to 1.0 eV to leave this site and jump to another anion [2].

Acceptor-doped BaZrO$_3$, particularly with high yttrium substitution, shows very high proton conductivities, clearly exceeding the oxide ion conductivities of the best oxide ion conductors below 1000 K [3]. BaZr$_{1-x}$Y$_x$O$_3$ is a very good mixed proton and oxygen ion conductor. At this time, it is not clear which material parameters influence or dominate the proton conductivity. Recent findings on whether crystal lattice size, defect chemistry and microstructure influence its proton diffusivity include: proton conductivity in BZY is not affected by the microstructure such as grains and grain boundaries [4]; the crystallite size in BZY may significantly alter the diffusion constant [5]; a large unit cell lattice parameter may favour the proton diffusivity significantly [6]. We compare in the present paper the diffusivities of protons in BaZr$_{0.9}$Y$_{0.1}$O$_3$ (BZY10) as obtained from impedance spectroscopy and quasi-elastic neutron scattering for various temperatures in order to distinguish grain boundary and bulk contributions.

## 2. Experimental

Stoichiometric amounts of BaCO$_3$, ZrO$_2$ and Y$_2$O$_3$ with 99% purity were ball milled in isopropanol for 48 hours, calcined for 10 hours at 1473 K, ball milled again for 48 hours and then calcined for 10 hours at 1673 K. The resulting powder was milled in a planetary mill and then uniaxially pressed at 10 MPa into prismatic bars for the neutron experiment (5 cm x 5 mm x 2 mm) and disks for impedance spectroscopy (2 mm thick, 10 mm diameter), then cold isostatically pressed at 200 MPa, and sintered for 24 hours at 1993 K. BZY is reported to have the P4mm structure with a = 4.2151 Å, and c = 4.2047 Å, c/a = 0.9975 [7]. The low resolution XRD data on our BZY10 showed good visual agreement with reference pattern JCPDS 01-089-2486 for Pm3m BaZrO$_3$ with a = 4.206 Å, and no obvious traces of second phase [8]. Discrepancies in structural data of our samples and literature data may be due to different synthesis conditions. Scherrer analysis of the peak widths revealed crystallite sizes not significantly exceeding 80 nm in any low indexed crystal orientation. The smallest width obtained was 45 nm. The theoretical density of the bar material is 6.20 gcm$^{-3}$. The actual density was 94% of this value. No further steps to increase the density and thus maximize the electrolyte performance were undertaken. Scanning electron microscopy (SEM) analysis showed that the grain size was around 2 μm and macropores were smaller than 0.7 μm.

The dry reference state for the QENS samples was obtained by keeping the samples for 14 hours at 1173 K under dry oxygen (p$_{H2O}$ ≤ 10 Pa). Humidification was achieved by passing wet oxygen (p$_{H2O}$ ≥ 2200 Pa) at 873 K for 24 hours. In the protonic regime, the formation of charge carriers and their mobility are independent of the atmosphere. [9]. Therefore, O$_2$ was chosen for simplicity. After drying (sample mass m$_d$) or humidifying (m$_h$), the samples were quenched and weighed with a high-precision microbalance to determine the proton content. The thus determined proton concentration in the samples prior to the QENS experiment, with M$_{BZY10}$ and M$_{H2O}$ being the respective molar weights, was 3 mol % and did not significantly decrease during the heating ramp.

$$[OH^{\cdot}] = \frac{m_h - m_d}{m_d} \times \frac{M_{BZY10}}{0.5 M_{H2O}} = 30.61 \cdot \frac{m_h - m_d}{m_d} \quad (2)$$

Conductivities were obtained by impedance spectroscopy (IS), with the sample temperature decreased from 1173 K to 300 K in steps of 50 K, in a ProboStat® cell (Norwegian Electro Ceramics AS). The samples had a disk shape of 15 mm diameter and 0.75 mm thickness. Electrical contact was provided via a platinum mesh on the top and bottom of the disk. A constant atmosphere was maintained by introducing oxygen humidified with water vapour up to $p_{H2O}$ ~ 2200 Pa. Spectra were recorded with a frequency sweep from 0.5 Hz to 1 MHz and oscillation amplitude of 1 V (Frequency Response Analyzer Solartron 1260) and fitted to a model circuit, representative of electrode area, grain boundary, and bulk. The conductivity σ was corrected for the sample geometry and translated into diffusivity D(T) via the Nernst-Einstein-Relation [10,11]:

$$\frac{\sigma}{D_H^0} = \frac{[OH^{\bullet}]}{V_m} \frac{eF}{kT} \exp\left(-\frac{E_H}{kT}\right) \quad (3)$$

[OH$^{\bullet}$] is the OH$^{\bullet}$ concentration, $V_m$ the molar volume, e the electron elementary charge and F the Faraday constant, $E_H$ the proton hopping energy in electron Volts and k the Boltzmann constant, and T the absolute temperature in K. The FOCUS time-of-flight instrument at the Swiss Spallation Neutron Source was used for quasi-elastic neutron scattering. The Q-range of 0.2 Å$^{-1}$ ≤ Q ≤ 0.85 Å$^{-1}$ was investigated with neutrons of 6 Å wavelength, corresponding to an incident energy of 2.273 meV. The instrument resolution was obtained by a separate measurement of a vanadium sample. Reduction of the raw data was performed with the DAVE software package [12]. The bars were arranged in a Pt cylinder so that the neutron beam passed one absorption length of the sample material and the container mounted in the heating chamber of the beamline. The QENS data obtained at 300 K served for background subtraction, assuming that protons are virtually immobile at this temperature. Spectra were then recorded with the temperature range 500 K ≤ T ≤ 900 K, in steps of 100 K. Data from 500 K and the data points at Q=0.2 Å$^{-1}$ and 0.3 Å$^{-1}$ at 600 K could not be analyzed due to poor signal to noise ratio.

## 3. Results and Discussion

A representative neutron scattering spectrum S(Q,ω) at 800 K is shown in Figure 1. All spectra were deconvoluted into an elastic (gaussian), quasielastic (lorentzian) and linear background scattering contribution, following a previously suggested procedure [2]. Proton motion in solids with diffusion constant D [cm$^2$s$^{-1}$] causes an inelastic neutron peak broadening Γ(Q) [meV] that simplifies for small Q [Å$^{-1}$] in least order approximation to [13]

$$\Gamma(Q) = 2\hbar D Q^2 \quad (4)$$

The full-width-at-half-maximum (FWHM) Γ [meV] of the Lorentzians that fitted our scattering curves are plotted vs. $Q^2$ in Figure 2 for different temperatures. Because of the linear trend of the FWHM vs. $Q^2$, we can conclude, in line with the Chudley-Elliott model, that the protons move by successive jumps, probably along oxygen ions in the lattice, as previously suggested by Karmonik and Hempelmann [14,15]. The slope of either linear curve corresponds to the 2ℏD, confirming that the diffusion constant D increases with increasing temperature. The Arrhenius representation of D(T) from QENS in Fig. 3 shows that the proton diffusion between 700 K and 900 K is thermally activated with an energy of $E_a$=0.13 eV, whereas below 700 K, the activation energy is only 0.04 eV. Since only two data points cover the range from 700 K – 900 K, the corresponding activation energy should not be taken too literally, but the trend of decreasing activation with increasing temperature is obvious.

Two ranges with different activation energies at around 700 K are also observed in other systems. For instance, the corresponding activation energies for $SrCe_{0.95}Yb_{0.05}O_{2.925}$ with 3-10 micrometer grain size are 0.61 eV and below 100 meV, respectively [15]. Data on nanocrystalline BZY with higher yttrium concentration and 7 nm crystallite size, $BaZr_{0.85}Y_{0.15}O_{2.925}$ (BZY15), indicate an activation energy of 0.17 eV for the high temperature range [5]. Thermal XRD studies on these samples suggest that substituting Zr by In, Ga, or Y inhibits crystallite growth significantly for temperatures above 1173 K. For T=1473 K, the Zr-substituted nanoparticles have crystallites of 20±3 nm, whereas the original barium zirconate exceeds 100 nm.

The thermal behaviour of our BZY10 is reflected by the impedance spectra. The semi-circles in the Nyquist plot, Fig. 4, are indicative of the grain boundary and bulk material resistance. The diffusion constant of our BZY10 as obtained by IS, differs by around two orders of magnitude and, not shown here, the material with the larger crystallites has the larger diffusion constant [6].

The activation energies obtained by QENS and IS cannot directly be compared because impedance spectroscopy generally probes the diffusion of protons and holes where present, whereas QENS, very sensitive to protons, probes the proton self-diffusion.

The activation energy for grain boundary and bulk conductivity as obtained by IS for our BZY10 is 1.21 eV and 0.46 eV for T ≤ 700 K, and 0.46 eV and 0.26 eV for T ≥ 700 K, respectively (Fig. 3). Bohn et al. [16] find $E_a$=0.44 eV for grain boundaries in BZY10. Poor grain boundary conductivity is also observed for $SrCe_{0.95}Yb_{0.05}O_{2.925}$ [2,14]. The activation energy for bulk diffusion in BZY was found to generally increase with increasing Y concentration [18]. Loss of Ba by evaporation during extended sintering may explain the relative enrichment of grain boundaries with $Y^{3+}$ dopant atoms [5], which presumably serve as traps for protons [14], and thus decreasing grain boundary conductivity [4]. This is supported by increased conductivities when excess BaO is used for synthesis [17]; $E_a$ = 0.34 eV for T ≤ 573 K, and 0.15 eV for T ≥ 573 K for grain boundaries. An alternative speculation is that the grain boundaries contain no Yttrium and, therefore, no $V_ö$. The activation energy for our BZY10 grain boundary conductivity, 0.46 eV, is close to that for bulk $BaZrO_3$, 0.42 eV [18]. Another interesting observation is that the activation energy from IS in BZY10 decreases significantly above 700 K by roughly a factor of 2 [17].

## 4. Conclusions

Two different diffusion processes cause two distinct dynamic ranges in the conductivities and diffusion as observed using IS. Two dynamic ranges are also visible in the QENS data at the same temperature of 700 K. We recall that QENS and IS may probe different processes. The fact that the activation energy and diffusion constants for T ≥ 700 K are in the same range for IS and QENS (Figs. 3) suggests that QENS resolves the proton diffusion process that is most efficient for electrolyte applications. It is another noteworthy observation that the diffusivities show two different slopes in the Arrhenius plot, suggesting two different activation energies for proton diffusion at different temperature ranges. Comparison with the work of Gross et al. [5] and Matzke et al. [2] confirms that a transition of modes of proton-phonon coupling takes place at 700 K.

# Acknowledgments

Financial support by the Swiss Federal Office of Energy, project # 100411, and by the European Commission, Marie Curie Actions contract # MIRG-CT-2006-042095 is gratefully acknowledged.

**Fig. 1** QENS peak (open symbols) with deconvolution in Gaussian (elastic) and lorentzian (quasielastic, dotted line) and least square fit I(Q,E) recorded at 900 K and Q = 0.75 Å$^{-1}$.

**Fig. 2** Full width at half maximum (FWHM) Γ of the quasielastic peak plotted against Q$^2$ reveals linear relation, suggesting that the Chudley-Elliott model is applicable.

**Fig. 3** Diffusion constant as obtained from QENS (open symbols) and from IS data. The IS data are shown for grain boundary (filled symbols) contribution and bulk (crosses).

**Fig. 4** Nyquist plots of BZY10 for 773 K, 673 K, and 573 K.

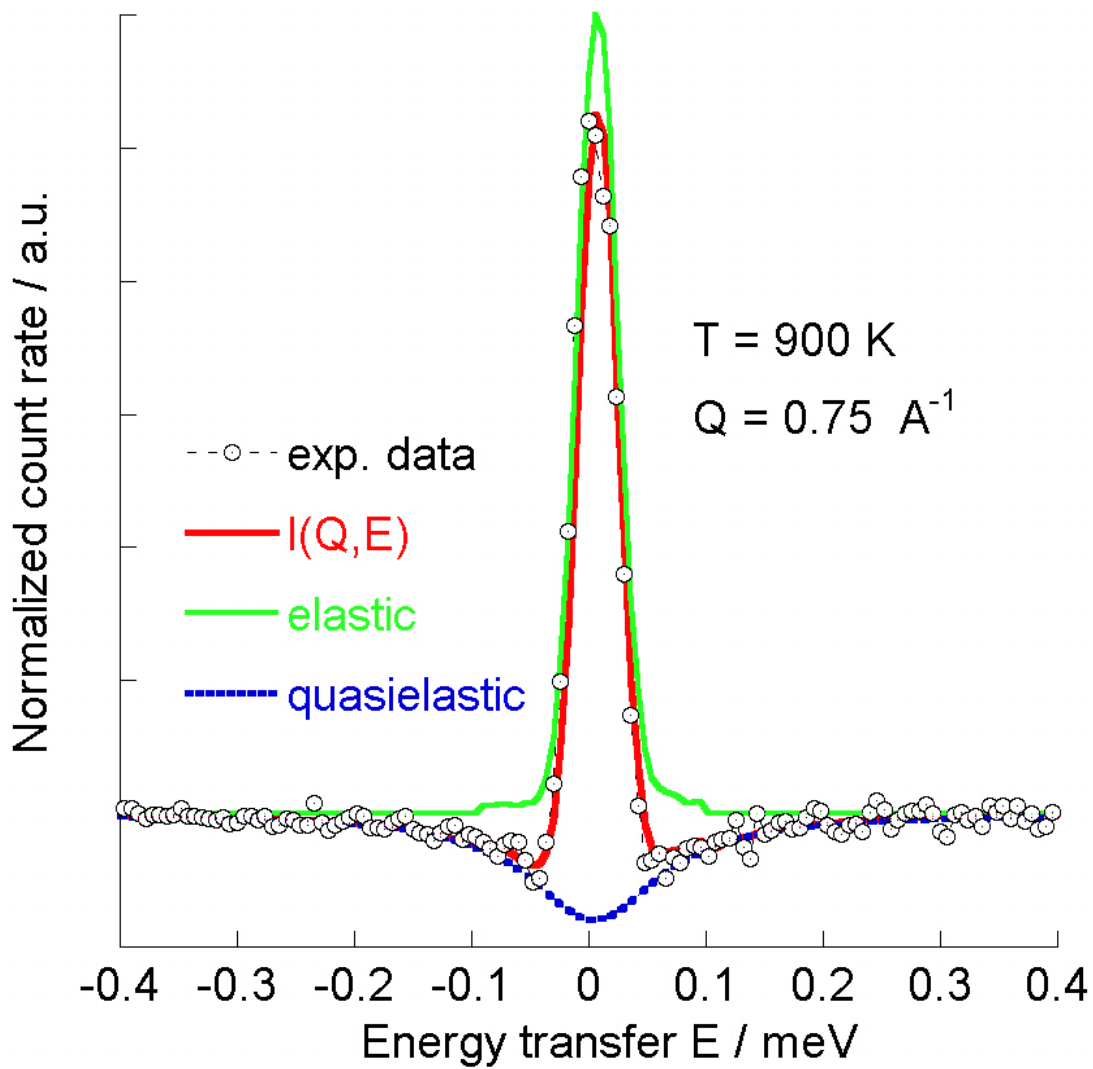

**Fig. 1**

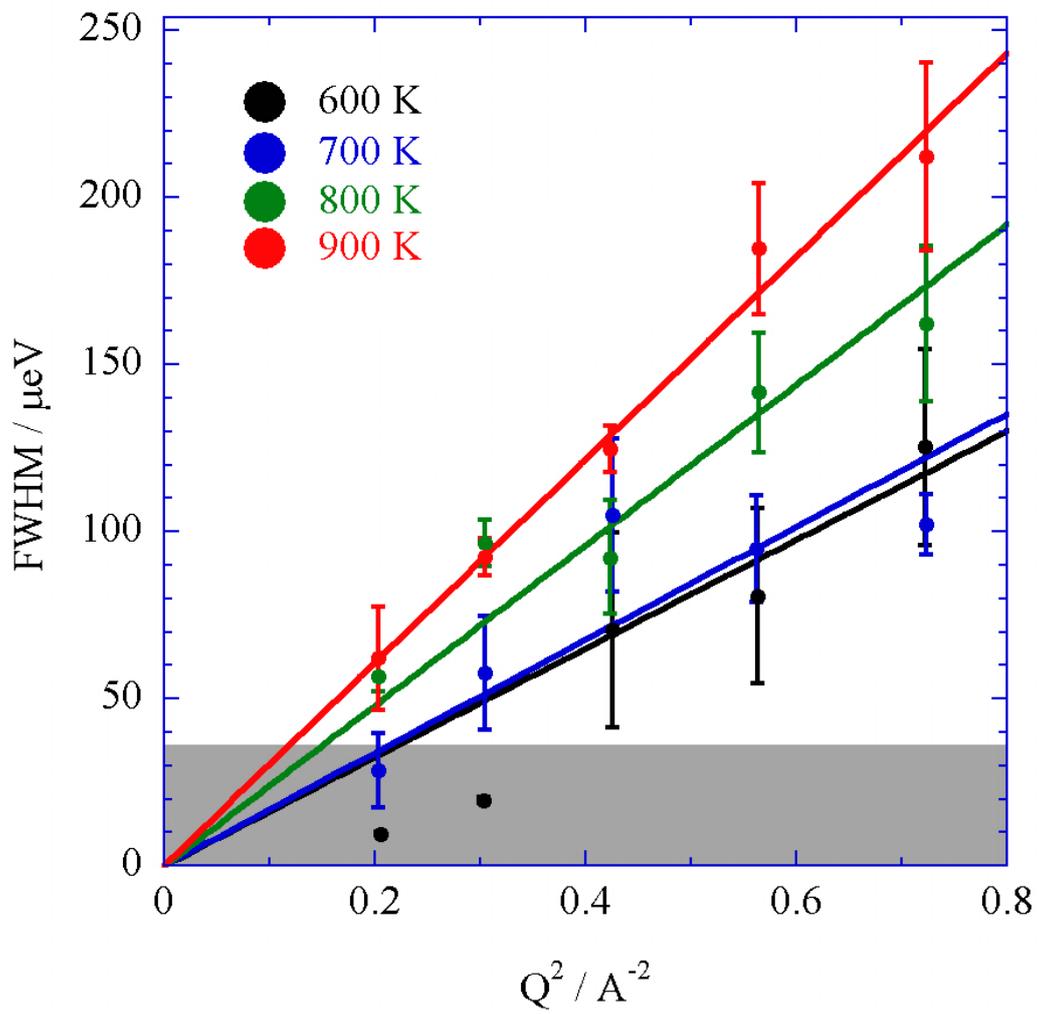

**Fig. 2**

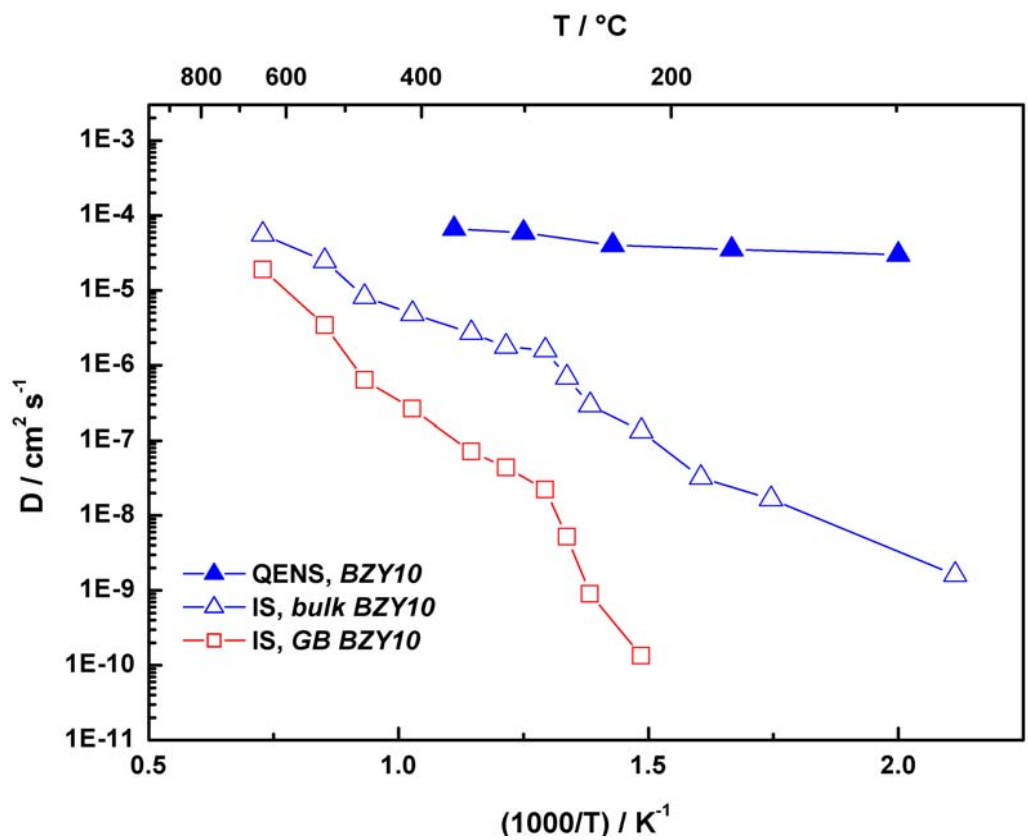

**Fig. 3**

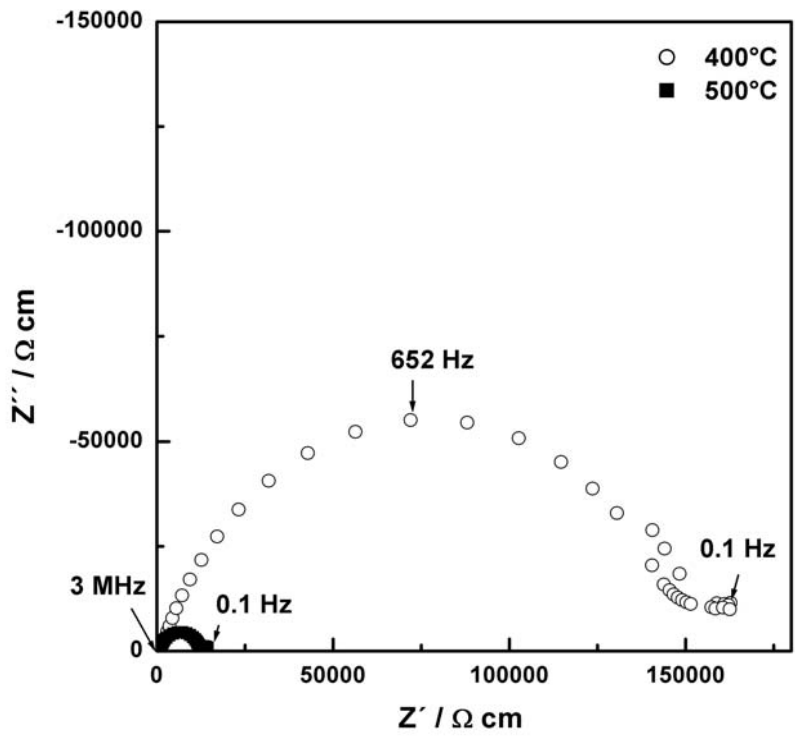

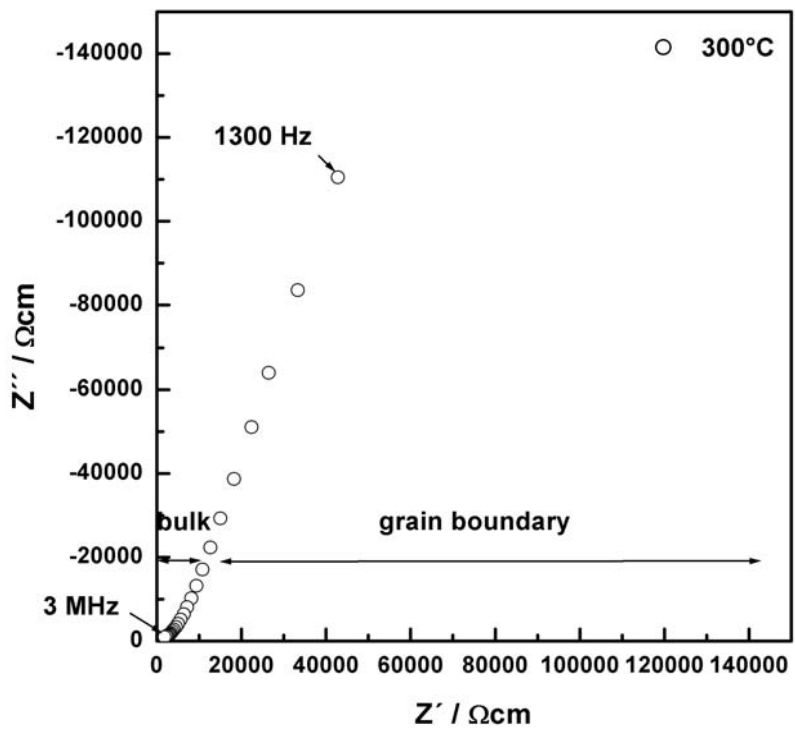

Fig. 4